# Universal self-field critical current for thin-film superconductors

E.F. Talantsev[1] & J.L. Tallon[1,2]

For any practical superconductor the magnitude of the critical current density, $J_c$, is crucially important. It sets the upper limit for current in the conductor. Usually $J_c$ falls rapidly with increasing external magnetic field, but even in zero external field the current flowing in the conductor generates a self-field that limits $J_c$. Here we show for thin films of thickness less than the London penetration depth, $\lambda$, this limiting $J_c$ adopts a universal value for all superconductors—metals, oxides, cuprates, pnictides, borocarbides and heavy Fermions. For type-I superconductors, it is $H_c/\lambda$ where $H_c$ is the thermodynamic critical field. But surprisingly for type-II superconductors, we find the self-field $J_c$ is $H_{c1}/\lambda$ where $H_{c1}$ is the lower critical field. $J_c$ is thus fundamentally determined and this provides a simple means to extract absolute values of $\lambda(T)$ and, from its temperature dependence, the symmetry and magnitude of the superconducting gap.

[1] Robinson Research Institute, Victoria University of Wellington, PO Box 33436, Lower Hutt 5046, New Zealand. [2] MacDiarmid Institute, Victoria University of Wellington, PO Box 33436, Lower Hutt 5046, New Zealand. Correspondence and requests for materials should be addressed to E.F.T. (email: Evgeny.Talantsev@vuw.ac.nz) or to J.L.T. (email: Jeff.Tallon@vuw.ac.nz).

Superconductors are characterized by two microscopic length scales: the London penetration depth, $\lambda$, and the coherence length, $\xi$. These strongly influence both their fundamental and applied behaviour, especially the critical current density $J_c$—above which the current becomes dissipative. In practical superconductors, which are all type II, $J_c$ is arguably its most important property and for example, in the high-$T_c$ superconductors, a huge effort has been expended in attempting to maximize $J_c$ as a function of temperature, $T$, and magnetic field, $H$[1]. $J_c$ falls rapidly with increasing external field, but even in zero external field the current flowing in the conductor generates a self-field that itself limits $J_c$. We refer to this limiting value as $J_c$(sf).

In a type-II superconductor, $J_c$ is widely thought to be governed by pinning of flux vortices as well as by geometrical factors arising from the detailed pinning microstructure. As a consequence, much of the above-noted effort has been applied to modifying and tuning this microstructure. On the other hand, nearly a century ago Silsbee[2] proposed that, for a type-I superconductor, the critical current 'is that at which the magnetic field due to the current itself is equal to the critical magnetic field'. In other words, the self-field $J_c$ is just that which is sufficient to generate a surface field equal to the critical field. By this was meant what we now understand to be the thermodynamic critical field, $H_c$, given by[3]

$$H_c(T) = \frac{\phi_0}{2\sqrt{2}\pi\mu_0\xi(T)\lambda(T)}, \quad (1)$$

where $\phi_0$ is the flux quantum and $\mu_0$ is the permeability of free space. Of course for a type-I superconductor, where flux vortices are absent, pinning is irrelevant and Silsbee's hypothesis is credible. $J_c$ may depend on geometry but not on microstructure. But for a type-II superconductor, the general consensus that pinning governs $J_c$ would insist that both geometry and microstructure are key players, and any kind of universal Silsbee criterion is untenable.

Here, by examining a wide range of experimental data, we ask whether this criterion does have any relevance to type-II superconductors. Surprisingly, the answer for conductors of thickness comparable to $\lambda$ is yes, and here the relevant critical field is the lower critical field $H_{c1}$, given by[3]

$$H_{c1}(T) = \frac{\phi_0}{4\pi\mu_0\lambda^2(T)}(\ln\kappa(T) + 0.5), \quad (2)$$

where $\kappa(T) = \lambda(T)/\xi(T)$ is the Ginsburg–Landau parameter, which is effectively constant under the logarithm. With this thickness constraint, we find for type-I superconductors:

$$J_c^I(\text{sf}) = \frac{H_c}{\lambda} = \frac{\phi_0\kappa(T)}{2\sqrt{2}\pi\mu_0\lambda^3(T)}, \quad (3)$$

and for type-II superconductors:

$$J_c^{II}(\text{sf}) = \frac{H_{c1}}{\lambda} = \frac{\phi_0}{4\pi\mu_0\lambda^3(T)}(\ln\kappa + 0.5). \quad (4)$$

As a consequence, $J_c$(sf) is fundamentally determined just by $\lambda$ and $\xi$ and independent of both geometry and microstructure. Because of the near constancy of $\ln(\kappa)$, for type-II superconductors $J_c$(sf) is dependent only on $\lambda$ and this then provides a simple means to extract absolute values of $\lambda(T)$ and, from its temperature dependence, the symmetry and magnitude of the superconducting gap. We present an indicative theoretical justification for this remarkably general and unexpected result, but we recognize that some questions remain to be resolved. We predict the doping and temperature dependence of $J_c$(sf) for $YBa_2Cu_3O_{7-\delta}$ (YBCO) as a test of our hypothesis. Hereafter, we consider only self-field $J_c$ values and therefore drop the identifier 'sf', except where we feel it is still needed.

## Results

**Basic model.** We consider a thin film of the type-II superconductor in the form of a long thin tape of rectangular cross-section in the $x$–$y$ plane and of thickness $2b$ and width $2a$, such that $b \ll a$. Our conductor is of quasi-infinite length along the $z$ axis in which a current of magnitude $I$ is flowing along its axis. The tape interior is defined by $-a \leq x \leq +a$ and $-b \leq y \leq +b$. According to London, in the Meissner state for small currents the self-field and transport current penetrate to a depth $\sim \lambda$, and the amplitude of the local surface current density, $J$, is[4]

$$J = \frac{B}{\mu_0\lambda} = \frac{H}{\lambda}, \quad (5)$$

where, in the usual notation, $B$ is the magnetic flux density and $H$ the field intensity, within the conductor surface. We use equation (5) to make an estimate of $J_c$ when $b \approx \lambda$. In this case (i) the current penetrates the entire cross-section, so that $J$ is no longer a surface current density but is approximately global across the film thickness, and (ii) when the $x$-component of $H$ reaches $H_{c1}$, vortices of opposite sign will tend to nucleate at the opposing surfaces and self-annihilate at the centre. They will do so both because of the Lorentz force driving them inwards and because of the attraction of overlapping vortices of opposite sign on opposite faces, which diverges logarithmically when $b < \lambda$. The consequent onset of dissipation defines $J = J_c$ where, from equation (5), $J_c \geq H_{c1}/\lambda$ and equality only applies if this force exceeds surface and bulk pinning forces. In the following, we observe that equality is indeed found for a wide range of superconducting materials and this leads immediately to equation (4).

This very approximate analysis leaves many open questions. For example, for $b > \lambda$ it is usual to discuss $J_c$ in terms of flux entry from the edges. This is discussed later. Our approach is simply to examine the available data from self-field $J_c$ studies on a wide variety of thin-film superconductors. If equation (4) does prove to be valid, then we have a simple means to determine absolute values of $\lambda$ (and the superfluid density $\rho_s \equiv \lambda^{-2}$) from measurements of $J_c(T)$. The test of success is how well inferred values of $\lambda$ concur with reported values. Moreover, the magnitude of the superconducting gap may then be determined from fitting the low-$T$ behaviour of $\rho_s$ as follows[5]. For $s$-wave symmetry:

$$\lambda(T)^{-2} = \lambda(0)^{-2}\left(1 - 2\sqrt{\frac{\pi\Delta(0)}{k_BT}}e^{-\Delta(0)/k_BT}\right), \quad (6)$$

while for $d$-wave symmetry:

$$\lambda(T)^{-2} = \lambda(0)^{-2}\left(1 - \sqrt{2}\frac{k_BT}{\Delta_m(0)}\right), \quad (7)$$

where $\Delta_m$ is the maximum amplitude of the $k$-dependent $d$-wave gap, $\Delta = \Delta_m\cos(2\theta)$.

**London penetration depth.** Figure 1 shows normalized plots of reported self-field $J_c(T)$ values (right-hand scale, arrowed) for a wide range of superconductors including type I, type II, $s$-wave and $d$-wave. Also plotted are the inferred values of $\lambda(T)$ (left-hand scale) calculated from the $J_c(T)$ values by inverting equations (3) or (4). Individual plots are presented and discussed in Supplementary Note 1. Panels (a) and (b) in Fig. 1 are $s$-wave, while (c) and (d) are $d$-wave superconductors. For both $s$- and $d$-wave cases, the $T$-dependence of $\kappa$, calculated from $\lambda \times \Delta$ (ref. 6), is weak and, in view of the logarithm and the cube root, we conveniently take $\kappa$ to be constant. The residual effect of a

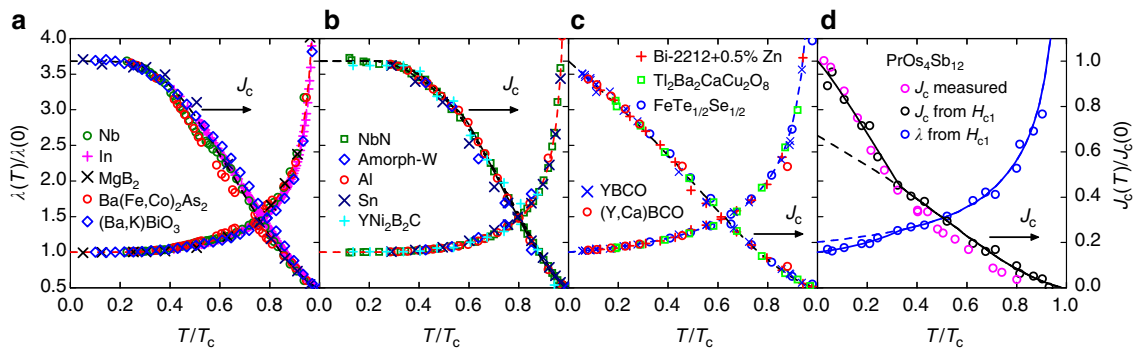

**Figure 1 | Summary of reported $J_c$ and calculated $\lambda$.** The temperature dependence of the normalized penetration depth, $\lambda(T)/\lambda(0)$ (left-hand scale) calculated from the normalized values of self-field critical current density, $J_c(T)/J_c(0)$ (right-hand scale, as indicated by arrows) for many different type-I and type-II superconductors. Values of $\lambda(T)$ are calculated using equations (3) and (4). (**a,b**) s-wave superconductors. (**c,d**) d-wave superconductors, as also seen by their very different low-T behaviour. The dashed red curves are the fitted s-wave weak-coupling T-dependence of $\lambda(T)$ and the dashed blue curves are the d-wave counterparts. The dashed black curves are $J_c$ back-calculated from these $\lambda(T)$ curves using equations (3) and (4). The deduced normalization parameters, $T_c$, $J_{c0}$ and $\lambda_0 = \lambda(0)$ are listed in Supplementary Table 1. (**d**) A slightly different analysis for $PrOs_4Sb_{12}$. Both $J_c$ and $\lambda$ are calculated from measurements of $H_{c1}$ using equation (4), and the calculated $J_c$ is compared with values measured from remnant magnetization (magenta data points). They are in excellent agreement. The two curves (dashed and solid) are obtained by adding two separate d-wave superfluid densities below 0.6 K.

T-dependent $\kappa$ is discussed in Supplementary Note 2. Values of $\kappa$ are sourced from the literature and are listed in Supplementary Table 1.

Our approach is as follows. From $J_c(T)$ data we calculate the $\lambda$ data points as plotted. Using $\lambda_0$ as the only fitting parameter, we then fit theoretical s-wave (dashed red) or d-wave (dashed blue) curves[6]. From these curves, we back-calculate $J_c(T)$ to give the dashed black curves, from which $J_{c0}$ is found. These deduced values of $\lambda_0$ and $J_{c0}$ are listed in Supplementary Table 1. The calculated $\lambda(T)$ data points are then fitted at low-T using equations (6) or (7) to determine $\Delta_0$. This is done using the nonlinear curve fit routine in the plot package 'Origin'. All data sources, film thicknesses and results are also summarized in Supplementary Table 1 along with reported values of $\lambda_0$ and $\Delta_0$ for comparison with our inferred values.

In Fig. 1, the radical difference between s-wave and d-wave symmetry at low T is immediately apparent. In the former case $J_c(T)$ exhibits an exponential plateau due to the isotropic gap, while in the latter $J_c(T)$ remains linearly increasing due to the nodal d-wave gap, in either case consistent with equations (6) or (7). Figure 1a shows examples of s-wave superconductors: Nb, In, $MgB_2$, $Ba(Fe,Co)_2As_2$ and $(Ba,K)BiO_3$. The fit with the weak-coupling s-wave model is excellent though $MgB_2$ and $Ba(Fe,Co)_2As_2$ show small deviations, possibly due to multiple gaps on distinct bands[7].

Figure 1b shows the same analysis for five samples where the fit is better using the dirty s-wave model. $YNi_2B_2C$ will be discussed later. For the type-I elements Al, Sn and In in Fig. 1, we have used equation (3). This is the so-called London depairing current density. We discuss this in relation to the Ginzburg–Landau depairing current density in Supplementary Note 5. For weak-coupling, s-wave superconductors $\kappa^{1/3}$ changes by <5% between $0 \le T \le T_c$ so by assuming constancy of $\kappa$ we again infer that $\lambda \propto 1/\sqrt[3]{J_c}$, but here with a different prefactor. The data for Al, Sn and In in Fig. 1a,b strongly support this analysis, and the deduced values of $\lambda_0$ shown in Supplementary Table 1 are in excellent agreement with directly measured values.

Next, Fig. 1c shows five d-wave examples: $FeTe_{0.5}Se_{0.5}$, $YBa_2Cu_3O_7$, 1% Ca-doped $YBa_2Cu_3O_7$, 0.5% Zn-doped $Bi_2Sr_2CaCu_2O_8$ and $Tl_2Ba_2CaCu_2O_8$. For each of these samples, the fit to the weak-coupling d-wave model is excellent across the entire temperature range. This is surprising because other techniques such as muon spin relaxation[8] suggest that the T-dependence of $\rho_s$ does not always follow the canonical d-wave form.

Finally, for Fig. 1d $PrOs_4Sb_{12}$, we used a different approach. From reported data for $H_{c1}$ (ref. 9), we calculated both $J_c(T)$ and $\lambda(T)$ from equation (4) using $\kappa = 29.7$ (refs 10,11). Both parameters reveal a transition to a second phase below 0.6 K (ref. 9), which results in an additional reduction of $\lambda(T)$. The two distinct curves (dashed and solid) are obtained by adding two separate d-wave superfluid densities below 0.6 K. Cichorek et al.[9] also determined $J_c(T)$ from remnant magnetization measurements. This is shown by the magenta symbols and, significantly, $J_c(T)$ shows an enhancement below 0.65 K that almost exactly mirrors our $J_c(T)$ values calculated from $H_{c1}$. This represents a quite exacting test of our central thesis.

All our results for calculated absolute values of $J_{c0}$, $\lambda_0$ and $\Delta_0$ are plotted versus measured values in Fig. 2. The error bars reflect the $2\sigma$-uncertainties in measured values of $\lambda_0$ or $\Delta_0$ summarized in Supplementary Table 1. In each case the correlation is excellent and it is this that validates our primary conclusion. Supplementary Note 3 shows how these results validate Silsbee's hypothesis. Figure 2a shows that, using equations (3) and (4), $J_c$ scales with $\lambda^{-3}$ over nearly three orders of magnitude. The only significant outlier is $YNi_2B_2C$ (ref. 12), but this is our only example for which $b > \lambda$. Applying the thickness-correction factor $(\lambda/b)\tanh(b/\lambda)$, as introduced below, the calculated $J_{c0}$ now falls close to the dashed line as indicated by the curved blue arrow.

Figure 2b shows values of $\lambda_0$ calculated from $J_c$. As hypothesized, the best $J_c$ values give values of $\lambda_0$ that match the measured values. In other films, where $J_c$ is low due to impurities, weak links or misalignment, $\lambda_0$ always exceeds the measured values. This is especially notable where a system has been improved over time, as for example, with the six films shown for $MgB_2$ (magenta data points). Here the inferred values of $\lambda$ descend towards the dashed line, as shown by the magenta arrow, as films were progressively improved. They do not fall below the line. A similar data progression over time for YBCO is discussed in Supplementary Note 6. $J_c(T)$ is thus fundamentally limited by $\lambda$ and not, for example, by the pinning, as we discuss further below. It is also important to recognize that the close correlation seen in Fig. 2b does not artificially arise simply from error reduction by taking the cube root to extract $\lambda$. This correlation is also seen without the cube root in Fig. 2a over now a much wider range, and also notably, where the relative order of type-I and type-II materials is significantly altered, but without loss of correlation.

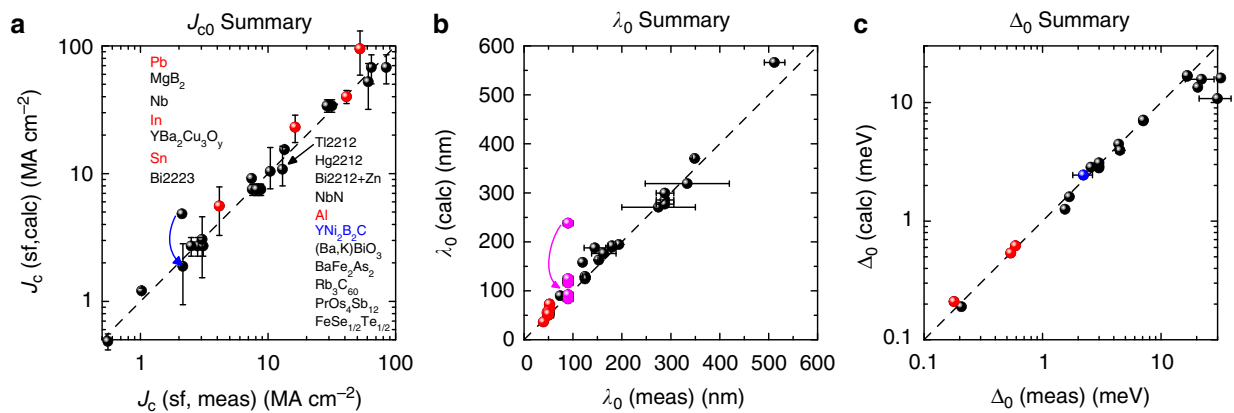

**Figure 2 | Summary comparison of calculated and measured values.** (**a**) Comparison of calculated thin-film self-field $J_{c0}$ values with measured $J_{c0}$ values. $J_{c0}$ is calculated from reported $\lambda$ values using equations (3) and (4). Red data points are type-I superconductors, black data points are type II. The blue arrow for YNi$_2$B$_2$C shows the effect of the correction factor $(\lambda/b)\tanh(b/\lambda)$, discussed in the text, when $b > \lambda$. Error bars reflect the range of reported $\lambda_0$ values—see Supplementary Table 1. The sample annotation follows the vertical order of the data points. (**b**) $\lambda_0$ values calculated from reported $J_c(T)$ data and plotted versus independently measured $\lambda_0$ values. The magenta arrow and symbols show the effect of improvement over time in self-field $J_c$ for six films of MgB$_2$. The data terminates on the dashed line where $J_c$ is now fundamentally limited by the superfluid density. Comparative data for YBCO over time is shown in Supplementary Figure 4. (**c**) Summary of values of $\Delta_0$ calculated from the low-$T$ behaviour of $\lambda(T)$ using equations (6) and (7), plotted versus measured values of $\Delta_0$.

**Energy gaps and symmetry.** We select two examples to illustrate the calculation of $\Delta_0$. Figure 3a,b shows the contrasting low-$T$ data for $\lambda(T)$ calculated from $J_c(T, sf)$ for (a) YBa$_2$Cu$_3$O$_7$ (this work) representing the $d$-wave case, and (b) five films of NbN (refs 13–15), representing the $s$-wave case. The dashed black curves are the data fits to equations (7) and (6), respectively. For YBa$_2$Cu$_3$O$_7$, the inferred gap value, $\Delta_0 = 16.6$ meV, compares well with the tunnelling measurements of Dagan et al.[16] (16.7 meV) and gives the ratio $2\Delta_0/k_BT_c = 4.26$ close to the $d$-wave weak-coupling ratio 4.28. It also compares well with the estimate of 17.7 meV from the condensation free energy[17], but infrared ellipsometry measurements give a higher value of 25 meV (ref. 18). And generally, our calculated values of $\Delta_0$ for the cuprate superconductors tend to be lower than reported values. Partly, this is due to the lack of very low-$T$ data for $J_c$, but in fact experimental values of gap magnitudes in the cuprates remain contentious. Tunnelling and ARPES data tend to show the presence of the (generally) larger pseudogap, and in our view the most reliable means of distinguishing the two is Raman scattering where $B_{2g}$ symmetry exposes the superconducting gap around the nodes, while $B_{1g}$ exposes the pseudogap around the antinodes[19]. In the present case, any sample that is optimally doped will have the pseudogap present and this will steepen the slope in $\lambda(T)$, thus reducing the inferred gap magnitude. The red crosses show the low-$T$ penetration depth measurements of Hardy et al.[20] and the agreement with ours, determined from $J_c$, is excellent.

For NbN, we show in Fig. 3b fits to five data sets[13–15]. Here $\langle\lambda_0\rangle$ is found to be 189 nm with $<3\%$ variation. The $s$-wave fits yield $\langle\Delta_0\rangle = 2.95$ meV with $<5\%$ variation and we find $2\Delta_0/k_BT_c = 4.13$, somewhat more than the weak-coupling $s$-wave limit of 3.53. Independent measurements[21] give $2\Delta_0/k_BT_c = 4.24$, which is rather close to our value.

Figure 2c shows inferred $\Delta_0$ values calculated for all 17 superconductors plotted against measured values as listed in Supplementary Table 1. There is generally good agreement over two orders of magnitude and across all systems. We note that equations (6) and (7) are restricted to the weak-coupling limit and this does not apply to all the samples investigated. In the case of Bi$_2$Sr$_2$Ca$_2$Cu$_3$O$_{10}$, the estimated value of $\Delta_0$ is particularly low and this is due to the presence of two-layer intergrowths as is evident in the original paper. This causes $\lambda(T)^{-2}$ to rise more rapidly below 90 K and thus yield a low value for $\Delta_0$. As noted, for the cuprates in general $\Delta_0$ values do tend to be low. This could be an indication of strong coupling but, if the samples are optimally doped then already the competing pseudogap is present[22] and this will diminish the inferred $\Delta_0$ values. This can only be clarified via the doping dependence of the low-$T$ behaviour of $J_c(T)$ where, in the sufficiently overdoped region, the pseudogap is no longer present.

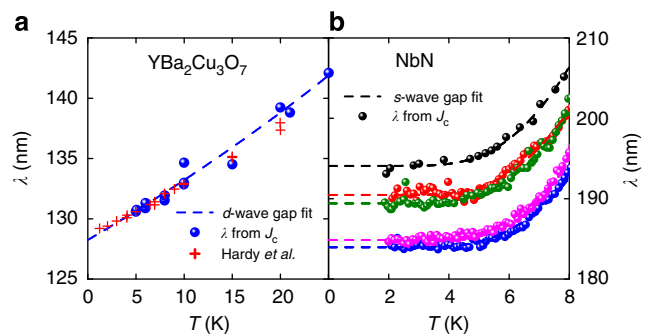

**Figure 3 | Low-temperature fits: gap symmetry.** (**a**) The low-$T$ fit to $\lambda(T)$ data determined from our measurements of $J_c(sf)$ for YBa$_2$Cu$_3$O$_7$ using equation (7). The fit yields $\lambda = 128.3$ nm and $\Delta_0 = 16.6$ meV. The red crosses show the low-$T$ penetration depth measurements of Hardy et al.[20] for comparison. (**b**) The low-$T$ fit to $\lambda(T)$ for NbN using equation (6) to determine $\Delta_0$. The characteristic flat $T$-dependence of $s$-wave superconductors at low $T$ is evident. The fits yield $\lambda = 189$ nm and $\Delta_0 = 2.95$ meV.

**Doping dependence of $J_c$ in YBa$_2$Cu$_3$O$_{7-\delta}$.** We conclude that the self-field $J_c(T)$ for high-quality, weak-link-free thin films with $b \leq \lambda$ appears to be a fundamental quantity, governed only by the absolute value of the superfluid density. If so we may use the superfluid density to predict the evolution of $J_c(T, p)$ with doping, $p$, for high-$T_c$ cuprates. Figure 4 shows the self-field $J_c$ calculated in this way from the ground-state doping-dependent superfluid density, $\rho_s(0)$, reported previously for $Y_{0.8}Ca_{0.2}Ba_2Cu_3O_{7-\delta}$ (refs 23,24). Using $\lambda(0) = 1/\sqrt{\rho_s(0)}$ and the theoretical $d$-wave $T$-dependence of $\rho_s(T)/\rho_s(0)$ (ref. 6), we generate the full $p$- and $T$-dependence of $J_c$, which is shown as a false-colour map in the $p$–$T$ plane in Fig. 4.

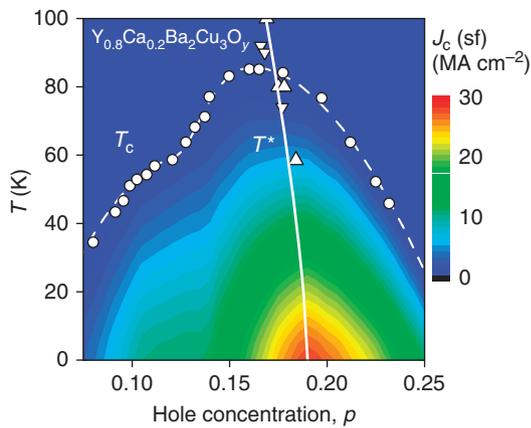

**Figure 4 | Predicted $J_c$ across the YBCO phase diagram.** Map of $J_c$(sf) across the phase diagram for $Y_{0.8}Ca_{0.2}Ba_2Cu_3O_y$ calculated from the superfluid density[23,24] using equation (4). A sharp peak is centred on the critical doping where the pseudogap $T^*$ line falls to zero (solid white curve). The triangles show the low-$T$ data points for $T^*$ reported from field-dependent resistivity[29]. A second smaller peak is predicted just below $P \approx 0.12$ where charge ordering has been reported[33]. The circles are $T_c$ data points.

Predicted values are seen to rise to a sharp maximum of about $30\,\text{MA cm}^{-2}$ centred at $p = 0.19$ holes per Cu in the slightly overdoped region, beyond optimal doping ($P \approx 0.16$) where $T_c$ reaches its maximum. (Doping with Ca introduces some impurity scattering that lowers the superfluid density. As a consequence, the maximum $J_c$ is less than that predicted from the superfluid density for Ca-free $YBa_2Cu_3O_7$: $37\,\text{MA cm}^{-2}$ for nearly fully oxygenated chains, and $\sim 42\,\text{MA cm}^{-2}$ for fully ordered chains[25].)

It is important to understand the significance of this skewed behaviour of $J_c(p)$ relative to optimal doping shown in Fig. 4. High-$T_c$ cuprates are characterized by the opening of a gap in the normal-state excitation spectrum, which is probably associated with reconstruction of the Fermi surface due to short-range magnetic order[26–28]. This phenomenon is associated with the so-called pseudogap that dominates the properties of optimally doped and under-doped cuprates, resulting in 'weak superconductivity' as indicated by a reduction in condensation energy, superfluid density and their associated critical fields[22].

The link between $J_c$ and the pseudogap is made by plotting in Fig. 4 the previously determined $T^*$ line where the pseudogap closes, as determined from field-dependent resistivity studies on epitaxial thin films[29,30]. The $T^*$ data extend to much higher temperatures, but some of the lower-$T$ data points are visible in the plot. The key result here is that $J_c$ maximizes just at the point where the pseudogap closes and $T^* \to 0$. Note, also, how the ridge in $J_c(T,p)$ follows the $T^*$ line, inclining towards lower doping at higher $T$ notwithstanding the fact that these two quantities are determined by quite different techniques. Clearly the rapid decline in $J_c$ below $T^*$ is due to the opening of the pseudogap and the consequent crossover to weak superconductivity. One is also impressed by the resemblance between this phenomenology and that associated with the presence of a quantum critical point[31], where a 'bubble' of high $J_c$ is centred on the point where $T^* \to 0$.

This sharp peak in $J_c(p)$ at $p = p_{\text{crit}}$ is recently confirmed by our wider group[32], but Fig. 4 also predicts a second smaller peak at $p \approx 0.12$. This second peak is also apparent in the upper critical field $H_{c2}$ of $YBa_2Cu_3O_{7-\delta}$ (ref. 33), and the search for a second peak in $J_c$ provides a strong test for the present ideas. A similar double peak in $\rho_s(p)$ is found in $La_{2-x}Sr_xCuO_4$ (ref. 34), suggesting that a double peak in $J_c(p)$ will be a common, perhaps universal, cuprate behaviour. Another test is that if $J_c(T)$ varies as $\rho_s^{3/2}$, then $J_c(0)^{2/3}$ should be diminished by impurity scattering in the same canonical way that the superfluid density is reduced[35]. This is distinguished by a much more rapid reduction in $\rho_s$ than in $T_c$. These ideas can be tested in Zn-substituted $YBa_2Cu_3O_{7-\delta}$ and have recently been confirmed by us.

**Refined model.** We return now to better justify the theoretical basis for our observations. The usual approaches to field distribution and critical currents are those of Brojeny and Clem[36] and Brandt and Indenbom[37], where for a very thin film the $y$-component of this field, $B_y(x = \pm a)$, diverges at the film edges[36]. As a consequence, Abrikosov vortices must enter from the edges and there then exists a domain extending in from the edges where the local current density $J(x)$ is constant $(=J_c)$ and in which these vortices are pinned. At the inner edge of this domain, $B_y$ falls to zero with infinite slope[37]. The onset of dissipation, defining the overall film $J_c$, occurs when this domain extends to the centre of the film at $x = 0$ and $J_c$ is now the global value not just the local value.

But, as we have suggested, an alternative approach is to consider the entry of vortices, not from the edges but from the large flat surfaces. For a very thin conductor the $x$-component of this self-field at the surface, $B_x(x, y = \pm b)$, is uniform across the width and of magnitude $\mu_0 b J$ (ref. 36). Consequently, if the current $I$ is increased until this field reaches $B = B_{c1}$, then vortices will nucleate at the flat surface in the form of closed loops around the conductor surface normal to the transport current[38]. These loops will tend to collapse inwards under the self-imposed Lorentz force ($\mathbf{J} \times \mathbf{B}$) on each vortex. It is only surface and bulk pinning that will prevent them from migrating to the conductor centre and self-annihilating there. Thus, inverting the above relation

$$J_c \geq B_{c1}/(\mu_0 b) = H_{c1}/b. \quad (8)$$

where the inequality arises from the, as yet, indeterminate role of pinning.

Now consider the interesting case when $b \approx \lambda$. These vortex loops now experience the additional attractive force of adjacent vortices of opposite sense located just $\lambda$ apart. This force becomes unbounded and thus inevitably overcomes pinning. The vortices mutually annihilate at the centre and the process continues indefinitely causing dissipation. This vortex entry from the faces defines a first critical current density given by the equality sign in equation (8), which is alternative to a second, which is associated with vortex entry from the edges. Which of these has the smaller $J_c$ and therefore is the operative mechanism?

To answer this, let us suppose that $J = H_{c1}/b$ over the full conductor cross-section and we calculate the degree of flux entry at the edges. Rearranging equation (13) of Brojeny and Clem[36] (or indeed equation (8) for a non-vanishingly-thin film) the $y$-component of the field near the edges for a uniform current distribution becomes

$$B_y(x) = \frac{B_{c1}}{2\pi} \ln\left[\left(\frac{x+a}{x-a}\right)^2\right]. \quad (9)$$

From this we find that $B_y$ falls to the value of $B_{c1}$ when $x = \pm 0.917a$, that is, perpendicular vortices can enter at either edge to only 4.2% of the film width. Thus the dominant mechanism for $J_c$ when $b \approx \lambda$ is flux entry from the large flat surfaces and $J_c$ is given by:

$$J_c(\text{sf}) = H_{c1}/b. \qquad (10)$$

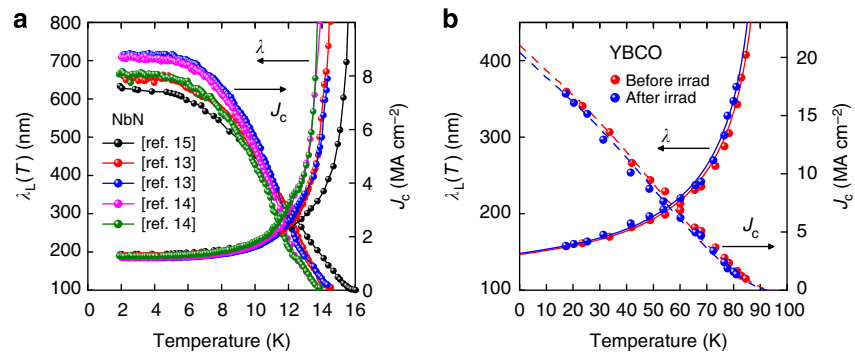

**Figure 5 | Thickness and irradiation dependence of $J_c$.** The $T$-dependence of $J_c$(sf) and $\lambda$ calculated from equation (4) for thin films of (**a**) NbN with different thicknesses, $b = 4$ nm (refs 13,14) and 11 nm (ref. 15) and different bridge widths. (**b**) YBCO before and after 'nano-dot' irradiation[45]. The solid red and blue curves are weak-coupling $d$-wave fits to the calculated $\lambda(T)$ data, before and after irradiation, respectively. The red and blue dashed curves are the respective back-calculated $J_c(T)$ curves from these $d$-wave fits. Note in **a** the small variation in $\lambda_0$ despite the quite large variation in $J_c$, and $\lambda_0$ is evidently independent of $b$ when $b < \lambda$. Values of $\Delta_0 = 2.95 \pm 0.02$ meV obtained from fitting the low-$T$ behavior of $\lambda(T)$ also reveal little variation. In **b**, $J_c$(sf) and $\lambda$ are independent of irradiation despite the large increase in pinning evidenced by a 60% increase in $J_c(T, H)$ above 1 Tesla.

We note that, when $b = \lambda$, equation (10) becomes equivalent to equation (4), but there is yet one final ingredient to add. The energy of formation of a vortex/anti-vortex pair on opposite faces of the film is reduced by an interaction term $\phi_0^2/(2\pi\mu_0\lambda^2) \times K_0(2b/\lambda)$ (ref. 38). Here $K_0(x)$ is the zeroth-order Bessel function of the second kind, and $2b$ is the vortex separation. This diverges logarithmically at small $b$ so that $B_{c1}$ is reduced as $B_{c1} \approx B_{c1}^\infty \times (b/\lambda)$, where $B_{c1}^\infty$ is the bulk value at large $b$. We adopt a heuristic crossover between the two limits in the form $B_{c1} = B_{c1}^\infty \tanh(b/\lambda)$. Combining with equation (10) and dropping the $\infty$ sign, we obtain our final result:

$$J_c(\text{sf}) = \frac{H_{c1}}{\lambda} \times (\lambda/b)\tanh\left(\frac{b}{\lambda}\right). \qquad (11)$$

where $H_{c1}$ is the bulk value. This concurs with equation (4) but with the additional correction factor $(\lambda/b)\tanh(b/\lambda)$. This accounts for all isotropic superconductors in Figs 1 and 2 with $b < \lambda$. In the case of anisotropic superconductors with $\lambda_x < \lambda_y$ (as in the case of high-$T_c$ cuprates), we simply rescale the problem $b \to b \times (\lambda_x/\lambda_y)$ and $J_c$ in equation (11) becomes:

$$J_c(T, \text{sf}) = \frac{H_{c1}(T)}{\lambda_{ab}(T)} \times (\lambda_c(T)/b)\tanh\left(\frac{b}{\lambda_c(T)}\right). \qquad (12)$$

Here we have replaced $\lambda_y$ by $\lambda_c$ and $\lambda_x$ by $\lambda_{ab}$, as is the usual convention for the cuprates, where $\lambda_c > \lambda_{ab}$. Equation (12) is the full generalization of equation (4). For $b < \lambda_c$, we recover equation (4), while for $b > \lambda_c$ we recover the $1/b$ falloff in $J_c$. A consequence of this is that if $b \leq \lambda_c(0)$, then equation (4) remains applicable to the highest temperatures because $\lambda_c(T)$ diverges as $T \to T_c$ and the small-$b$ limit is preserved. Let us now compare this functional dependence on $b$ with experimental data.

Figure 5a shows our analysis using equation (4) of five sets of $J_c(T)$ data for NbN. Four[13,14] have $b = 4$ nm and one[15] has $b = 11$ nm, all much less than $\lambda = 194$ nm. Despite these different thicknesses, all five data sets yield values of $\lambda$ close to this measured value (see Supplementary Table 1). The implication is that equation (4) is accurate for all $b \leq \lambda$. In contrast, for the alternative case $b > \lambda$, Stejic et al.[39] report for Nb-Ti films an inverse thickness dependence of $J_c(T)$, where $J_c \propto 1/b$. This is precisely what we argue above.

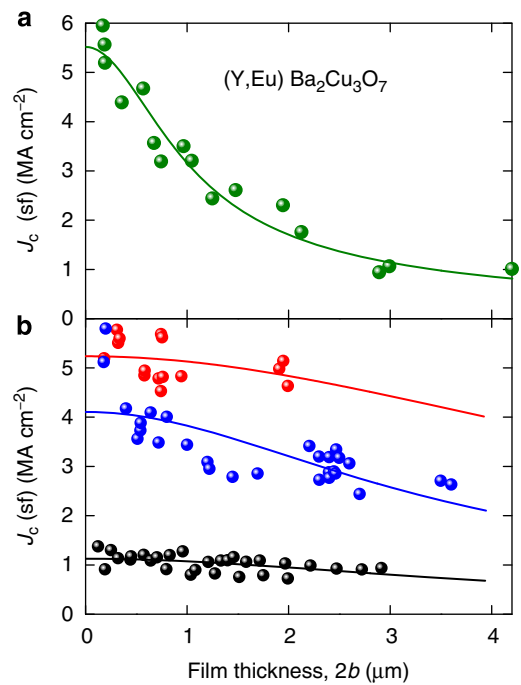

**Figure 6 | Thickness dependence of $J_c$ for YBCO.** $J_c$(sf) for epitaxial YBCO films versus film thickness. (**a**) $(Y_{0.67}Eu_{0.33})Ba_2Cu_3O_{7-\delta}$ at $T = 75.5$ K, from Zhou et al.[40] The solid curve is equation (12) with $\lambda_c = 620$ nm. This value is too low and reflects increasingly non-uniform composition and microstructure across the thickness. (**b**) YBCO systems with a highly uniform through-thickness microstructure: Feldmann et al.[42] (black data points and curve), with $\lambda_{ab}(77) = 339 \pm 5$ nm and $\lambda_c(77) = 1{,}153 \pm 238$ nm progressively thinned by ion milling; Zhou et al.[43] (blue data points and curve) gives $\lambda_{ab}(75.6) = 253 \pm 4$ nm and $\lambda_c(75.6) = 1{,}063 \pm 28$ nm; Feldmann et al.[44] (red data points and curve) gives $\lambda_{ab}(75.6) = 233 \pm 2$ nm and $\lambda_c(75.6) = 1{,}980 \pm 790$ nm.

Pursuing this further, Zhou et al.[40] report self-field $J_c$ measurements on many epitaxial thin films of $YBa_2Cu_3O_{7-\delta}$ and $(Y_{0.67}Eu_{0.33})Ba_2Cu_3O_{7-\delta}$ of varying thickness deposited on $SrTiO_3$ by pulsed laser deposition. Figure 6a shows the $J_c$ data measured at 75.5 K as a function of film thickness (data points). Also plotted is equation (12) fitted to this data—solid curve. The excellent agreement is misleading. The value of $\lambda_c = 620$ nm is too low and reflects a non-uniform composition and microstructure, which grows with increasing film thickness. Fits to other available data including Foltyn et al.[1] and Arendt et al.[41] give similar $\lambda_c$ values, 634 and 654 nm, respectively. Again these are smaller than the expected $\geq 1,100$ nm. In contrast, Fig. 6b shows data for films with an intentionally highly uniform through-thickness microstructure[42–44]. In the case of Feldmann[42], this uniformity was confirmed by progressive ion milling of a single film. Here the $\lambda_c$ values shown in the figure are now indeed realistic.

We conclude that equations (11) and (12) provide a good description of $J_c$ for uniform films in the general case when $b \neq \lambda$. Of course, as $b$ is increased, the alternative $J_c$ mechanism involving flux entry from the edges must eventually become dominant.

**Pinning**. Our claim is that self-field $J_c$ is independent of pinning when $b \leq \lambda$ unless, perhaps, very strong pinning is introduced. We illustrate this with one of the few examples available. Lin et al.[45] scribed an array of nanocolumns into a YBCO microbridge using a focused electron beam. The array had a lattice constant of 90 nm (corresponding to a matching field of 0.25 Tesla) and diameters about double the coherence length at 77 K, while the film thicknesses were 50 and 90 nm, thus satisfying our condition $b \leq \lambda$. We show our analysis of the $H = 0$ self-field $J_c$ in Fig. 5b. These authors report a 60% increase in the in-field $J_c(T)$ above 1 Tesla but, we note, there is essentially no change in $J_c(T, sf)$, indeed a small decrease consistent with a small loss of effective cross-sectional area. There is no change in the inferred $\lambda_0$. Thus the evident increase in bulk pinning and inevitable changes in surface roughness and surface pinning have no apparent effect on $J_c(T, sf)$, consistent with our hypothesis. Similarly, with 25 MeV $^{16}$O ion irradiation of YBCO films Roas et al.[46] report an 80% increase in $J_c$ above 1 Tesla, but a significant reduction in $J_c(sf)$. On the other hand, a later study by this group[47] showed fast neutron irradiation lifted $J_c(sf)$ at 4.2 K from 19.4 to 32 MA cm$^{-2}$. While this is no more than the $J_c(0, sf)$ values we quote above for our pristine films, it suggests that strong pinning could, in the extreme, overtake the Silsbee mechanism we advance here. But here they use an extremely high electric field criterion of 50 µV cm$^{-1}$, and the apparent enhancement could simply be caused by a reduction in $n$-value due to irradiation. Later neutron irradiation studies[48–50] on epitaxial YBCO films, reported only a detrimental effect on $J_c(sf)$ while lifting in-field performance. On the basis of this, we feel it still remains to be confirmed that pinning centres created by irradiation can improve $J_c(sf)$. Further literature examples supporting the pinning independence of $J_c(sf)$ are given in Supplementary Note 4.

## Discussion

We have shown that for thin films of thickness $b < \lambda$, the self-field $J_c$ is given by $H_c/\lambda$ for type-I superconductors and $H_{c1}/\lambda$ for type-II superconductors. This provides a simple, direct means to determine the absolute magnitude of the penetration depth and means that, contrary to widespread thinking, $J_c(T, sf)$ is a fundamental property independent of pinning landscape and microstructural architecture. We have thus confirmed Silsbee's hypothesis for all the superconductors we have examined. In the case of the cuprates, our prediction of a sharp peak in $J_c(p, sf)$ at the critical doping, where $T^\star \to 0$, is borne out in separate studies[32].

To conclude, we suggest the following possible tests of the ideas we have presented. We predict a second peak in $J_c(p, sf)$ near $p \approx 0.12$, near which charge ordering occurs[33], and we suggest that, for Zn-substituted $YBa_2Cu_3O_y$, $J_c(sf)^{2/3}$ will be suppressed by impurity scattering in the same canonical manner as the superfluid density. Along with this the $T$-dependence of $J_c(sf)$ should cross over from a linear-in-$T$ behaviour to $T^2$ consistent with the superfluid density[6]. A further key test would be to measure and correlate both superfluid density and self-field $J_c$ in a single film as a function of progressive disorder via irradiation. Inspection of equation (11) shows that, with increasing film thickness, $J_c(sf)$ should cross over from a $\lambda^{-3}$ dependence when $b < \lambda$ to a $\lambda^{-2}$ dependence when $b > \lambda$. It should be relatively straightforward to test this. The approach reported here should readily translate to superconducting nanowires and could be used to measure the transport mass anisotropy in layered superconductors by comparing $J_c(sf)$ measurements in $a$-axis- and $c$-axis-aligned films. A key challenge will be to treat the crossover from a Silsbee-dominated mechanism in self-field to a more conventional pinning-dominated mechanism with flux entry from the edges as external field is increased. The implications for a.c. loss should be explored and, finally, the model we present here lends itself to an error-function onset to resistance (rather than the conventional power law) due to the local distribution of superfluid density[51].

## Methods

**Data sources—self-field critical current density**. Of the vast literature for $J_c$, surprisingly few data sets are available that meet the collective requirements for the present analysis. These are as follow: we require transport (not magnetization) $J_c$ data; data reported under self-field conditions; $J_c$ data for weak-link-free thin films; in which $b \leq \lambda$; and that extend down to temperatures, $T \leq 0.2T_c$. The analyses reported here more or less exhaust the available data.

**Data sources—penetration depth**. We have chosen to test equations (3) and (4) using literature data for $\lambda_0$ and not for $H_c$ and $H_{c1}$ for which even recent literature shows quite divergent values. An illustrative example reports a breakdown of the Uemura relation between $T_c$ and $\rho_s$ by a factor of eight, based on $H_{c1}$ data for $Ba_{0.6}K_{0.4}Fe_2As_2$ (ref. 52). Subsequent measurements of superfluid density using muon spin relaxation showed that this system was in fact in full agreement with the Uemura relation[53–55]. Early penetration depth data are variable in quality, and microwave measurements often do not yield absolute values, probing as they do values of $\Delta\lambda = \lambda(T) - \lambda(0)$, only. Where possible we have relied on muon spin relaxation or polarized neutron reflectometry for $\lambda$ values, because these directly probe the field profile and tend to be rather reproducible from one group to another. Where possible we also give multiple sources, and the ranges from these sources are reflected in Supplementary Table 1 and the error bars in Fig. 2 and Supplementary Fig. 2.

## References

1. Foltyn, S. R. et al. Materials science challenges for high-temperature superconducting wire. *Nat. Mater.* **6,** 631–642 (2007).
2. Silsbee, F. B. Note on electrical conduction in metals at low temperatures. *J. Franklin Inst.* **184,** 111 (1917).
3. Poole, C. P., Farach, H. A., Creswick, R. J. & Prozorov, R. *Superconductivity* Chaps 2, 11, 12, 14 (Academic Press, 2007).
4. London, H. Phase-equilibrium of supraconductors in a magnetic field. *Proc. R. Soc. Lond. A* **152,** 650 (1935).
5. Prozorov, R. & Kogan, V. G. London penetration depth in iron-based superconductors. *Rep. Prog. Phys.* **74,** 124505 (2011).
6. Won, H. & Maki, K. d-wave superconductor as a model of high-$T_c$ superconductors. *Phys. Rev. B* **49,** 1397–1402 (1994).
7. Kogan, V. G., Martin, C. & Prozorov, R. Superfluid density and specific heat within a self-consistent scheme for a two-band superconductor. *Phys. Rev. B* **80,** 014507 (2009).
8. Sonier, J. E. et al. Hole-doping dependence of the magnetic penetration depth and vortex core size in $YBa_2Cu_3O_y$: evidence for stripe correlations near 1/8 hole doping. *Phys. Rev. B* **76,** 134518 (2007).
9. Cichorek, T. et al. Pronounced enhancement of the lower critical field and critical current deep in the superconducting state of $PrOs_4Sb_{12}$. *Phys. Rev. Lett.* **94,** 107002 (2005).
10. MacLaughlin, D. E. et al. Muon spin relaxation and isotropic pairing in superconducting $PrOs_4Sb_{12}$. *Phys. Rev. Lett.* **89,** 157001 (2002).
11. Bauer, E. D., Frederick, N. A., Ho, P.-C., Zapf, V. S. & Maple, M. B. Superconductivity and heavy fermion behavior in $PrOs_4Sb_{12}$. *Phys. Rev. B* **65,** 100506(R) (2002).

### Acknowledgements

J.L.T. thanks the Marsden Fund of New Zealand and the MacDiarmid Institute for Advanced Materials and Nanotechnology for financial support. We also thank N.M. Strickland, S.C. Wimbush and J.G. Storey for developing the HTS-based $J_c$ measuring rig, and Professor A.M. Campbell, S.C. Wimbush, A. Malozemoff and the referees for helpful comments on this manuscript.

### Author contributions

Both authors contributed equally to all aspects of this work.

### Additional information

**Supplementary Information** accompanies this paper at http://www.nature.com/naturecommunications

**Competing financial interests:** The authors declare no competing financial interests.

**Reprints and permission** information is available online at http://npg.nature.com/reprintsandpermissions/

**How to cite this article:** Talantsev, E. F. & Tallon, J. L. Universal self-field critical current for thin-film superconductors. *Nat. Commun.* 6:7820 doi: 10.1038/ncomms8820 (2015).